\documentclass{article}
\usepackage[english]{babel}
\usepackage[english]{babel}
\usepackage{graphicx,multicol}
\usepackage{epic,eepic,epsfig}
\usepackage{amssymb}
\usepackage{amsmath}
\usepackage{color}
\usepackage{epsfig,url}
\usepackage{multirow}
\newcommand{\SB}[1]{{#1}}

\usepackage{eso-pic} 

\newcommand{\induce}[2]{\mbox{$ #1 \langle #2 \rangle$}}

\newcommand{\dom}{\mbox{$\rightarrow$}}
\newcommand{\sdom}{\mbox{$\Rightarrow$}}
\newcommand{\qed}{\hfill$\diamond$\\\vspace{2mm}}
\newcommand{\pf}{{\bf Proof: }}
\newcommand{\be}{\begin{enumerate}}
\newcommand{\ee}{\end{enumerate}}
\newcommand{\bd}{\begin{description}}
\newcommand{\ed}{\end{description}}

\newcommand{\beq}{\begin{equation}}
\newcommand{\eeq}{\end{equation}}

\newtheorem{theorem}{Theorem}[section]
\newtheorem{lemma}[theorem]{Lemma}
\newtheorem{proposition}[theorem]{Proposition}

\newtheorem{corollary}[theorem]{Corollary}

\newtheorem{claim}{Claim}
\newtheorem{observation}{Observation}

\newcommand{\FF}{{\cal F}}

\begin{document}
\bibliographystyle{plain}

\title{Out-colourings of Digraphs} \author{N. Alon\thanks{Sackler
    School of Mathematics and Blavatnik School of Computer Science,
    Tel Aviv University, Tel Aviv, Israel, Email: nogaa@tau.ac.il.
    Research supported in part by a BSF grant, an ISF grant and a GIF
    grant.}\and J. Bang-Jensen\thanks{Department of Mathematics and
    Computer Science, University of Southern Denmark, Odense DK-5230,
    Denmark, Email: jbj@imada.sdu.dk. Financial support: Danish
    research council, grant number 1323-00178B.}\and
  S. Bessy\thanks{LIRMM, Universit\'e de Montpellier, Montpellier,
    France.  Email: stephane.bessy@lirmm.fr. Financial support: OSMO
    project, Occitanie regional council.}}

\maketitle

\begin{abstract}
We study vertex colourings of digraphs so that no out-neighbourhood is
monochromatic and call such a colouring an {\bf out-colouring}. The
problem of deciding whether a given digraph has an out-colouring with
only two colours (called a 2-out-colouring) is ${\cal
  NP}$-complete. We show that for every choice of positive integers
$r,k$ there exists a $k$-strong bipartite tournament which needs at
least $r$ colours in every out-colouring. Our main results are on
tournaments and semicomplete digraphs. We prove that, except for the
Paley tournament $P_7$, every strong semicomplete digraph of minimum
out-degree at least 3 has a 2-out-colouring. Furthermore, we show that
every semicomplete digraph on at least 7 vertices has a
2-out-colouring if and only if it has a {\bf balanced} such colouring,
that is, the difference between the number of vertices that receive
colour 1 and colour 2 is at most one. In the second half of the paper
we consider the generalization of 2-out-colourings to vertex
partitions $(V_1,V_2)$ of a digraph $D$ so that each of the three
digraphs induced by respectively, the vertices of $V_1$, the vertices
of $V_2$ and all arcs between $V_1$ and $V_2$ have minimum out-degree
$k$ for a prescribed integer $k\geq 1$.  Using probabilistic arguments
we prove that there exists an absolute positive constant $c$ so that
every semicomplete digraph of minimum out-degree at least
$2k+c\sqrt{k}$ has such a partition. This is tight up to the value of
$c$.
\end{abstract}

\section{Introduction}

A {\bf $\mathbf{k}$-partition} of a (di)graph $G=(V,A)$ is a partition
of $V$ into $k$ disjoint non-empty subsets $V_1,\ldots{},V_k$.  The
complexity of deciding whether one can partition the vertex set of a
(di)graph into two or more non-empty subsets of vertices, such that
the sub(di)graphs induced by these sets satisfies prescribed
properties is a difficult problem. In \cite{bangTCS640,bangTCS636} the
complexity of 120 such problems concerning 2-partitions of digraphs
was settled.

Thomassen \cite{thomassenC3} proved that every digraph with minimum
out-degree at least 3 has a 2-partition $(V_1,V_2)$ such that each of
the subdigraph induced by these two sets has minimum out-degree at
least 1, see also \cite{alonJCT68} for an extension for partitioning
into more parts. One can decide in polynomial time whether a
  given digraph has a 2-partition such that each set has minimum
  out-degree at least 1 (see e.g. \cite{bangTCS640}). Another result
by Thomassen on digraphs with no even directed cycle
\cite{thomassenEJC6} implies that there is no lower bound on the
minimum out-degree of a digraph that implies that a digraph $D$ has a
2-partition $(V_1,V_2)$ such that the bipartite digraph $D(V_1,V_2)$
induced by the arcs between $V_1,V_2$ has minimum out-degree at least
one. Still, as was shown in \cite{bangman17}, it can be decided in
polynomial time whether a given digraph has a 2-partition $V_1,V_2$
such that $D(V_1,V_2)$ has minimum out-degree at least one (if we
require higher out-degrees, the problem becomes ${\cal NP}$-complete
\cite{bangbipman17}).

In the language of 2-partitions of digraphs a 2-out-colouring is the
same thing as a 2-partition $(V_1,V_2)$ such that each vertex has an
out-neighbour in both sets. That is, we want both of the properties
above. This is the same thing as searching for a 2-colouring with no
monochromatic edges of the hypergraph ${\cal H}_D=(V,{\cal E})$ on
$|V|$ edges that we obtain from a given digraph $D=(V,A)$ by letting
the edges in $\cal E$ correspond to each of the $|V|$
out-neighbourhoods in $D$. It is well-known \cite{lovasz1973} that it
is ${\cal NP}$-complete to decide whether one can 2-colour the
vertices of a hypergraph ${\cal H}$ in such a way that no edge is
monochromatic. As we show below this problem remains ${\cal
  NP}$-complete for hypergraphs like ${\cal H}_D$ that are the {\bf
  out-neighbourhood hypergraph} of some digraph $D$.

Our main focus is on tournaments and semicomplete digraphs, that is,
respectively, orientations of complete graphs and digraphs with no
pair of non-adjacent vertices. Tournaments form the most well-studied
class of digraphs and despite their restricted structure a lot of deep
results exist and many challenging problems remain for this class (see
e.g. \cite{bangTchapter} for a comprehensive list of results on
tournaments and semicomplete digraphs). In this paper we show, among
other results, that with only one exception on 7 vertices, every
semicomplete digraph with minimum out-degree at least 3 has a
2-out-colouring and give a polynomial algorithm for finding such a
partition or certifying its nonexistence in an arbitrary semicomplete
digraph. In the second part of the paper we consider a generalization
of 2-out-colourings where we do not only want that each vertex has at
least one out-neighbour in both sets but now we want at least $k$
out-neighbours in each set for a prescribed number $k$. Using
probabilistic arguments we prove that every tournament whose minimum
out-degree is sufficiently large as a function of $k$ has such a
partition. The bound we give on this function is asymptotically best
possible.

The paper is organized as follows. We first provide some notation and
show that even for highly structured digraphs, such as symmetric
digraphs and bipartite tournaments, high out-degree is not sufficient
to guarantee the existence of an out-colouring with $r$ colours for
any fixed $r$. In Section~\ref{complexitysec} we show that deciding
whether a digraph has a 2-out-colouring is ${\cal NP}$-complete, even
when the input is restricted to be a symmetric digraph with high
out-degree or to be semicomplete bipartite. In
Section~\ref{Toutcolsec} we characterize all tournaments which have a
2-out-colouring. Using these results we characterize also in
Section~\ref{SDoutcolsec} all semicomplete digraphs admitting a
2-out-colouring. In Section~\ref{Balanced} we prove that with very few
exceptions, if a semicomplete digraph has a 2-out-colouring, then it
also has a balanced one, that is, the sizes of the two sides of the
2-partition differ by at most one. In Section~\ref{kinbothsets} we
study the generalization of 2-out-colourings where we want a
2-partition such that each vertex has at least $k$ out-neighbours in
both sets. Using probabilistic methods we give asymptotically best
possible bounds for the minimum out-degree which will guarantee that a
tournament with this minimum out-degree has a 2-partition as
above. Finally, in Section~\ref{remarksec} we mention some further
consequences of our results.

\section{Terminology and preliminaries}

Notation follows \cite{bang2009}.  For any digraph $D=(V,E)$ we use
the notation $u\rightarrow v$ if the arc $uv$ belongs to $E$ and we
say that $u$ {\bf dominates} $v$ and say that $v$ is an {\bf
  out-neighbour} of $u$ and $u$ is and {\bf in-neighbour} of $v$. The
set of out-neighbours (in-neighbours) of a vertex $v$ is denoted by
$N^+(v)$ ($N^-(v)$) and we denote by $\delta^+(D)$ the minimum
out-degree of the digraph $D$, that is the minimum size of an
out-neighbourhood.  For two disjoint sets of vertices $A$ and $B$ of
$D$ we write $A\Rightarrow B$ if $a\rightarrow b$ for every $a\in A$
and $b\in B$.  A set $X\subset V$ is an {\bf in-dominating set} in
$D=(V,A)$ if every vertex in $V-X$ dominates at least one vertex in
$X$.

A $(u,v)$-path is a directed path from $u$ to $v$.  A digraph is {\bf
  strongly connected} (or {\bf strong}) if it contains a $(u,v)$-path
for every ordered pair of distinct vertices $u,v$.  A digraph $D$ is
{\bf $k$-strong} if for every set $S$ of less than $k$ vertices the
digraph $D-S$ is strong.  A {\bf strong component} of a digraph $D$ is
a maximal subdigraph of $D$ which is strong. A strong component is
{\bf trivial}, if it has order $1$. An {\bf initial} (resp. {\bf
  terminal}) strong component of $D$ is a strong component $X$ with no
arcs entering (resp. leaving) $X$ in $D$. It is easy to see that for a
tournament $T$ which is not strong, we can order the strong components
uniquely as $D_1,\ldots{},D_k$, $k\geq 2$ so that all arcs are
directed from $D_i$ to $D_j$ for $1\leq i<j\leq k$. In particular $T$
has exactly one initial component $D_1$ and exactly one terminal
component $T_k$.

A {\bf $\mathbf{k}$-colouring} $\gamma$ of a digraph $D=(V,E)$ is a
mapping $\gamma:V\rightarrow \{1,\dots ,k\}$ of its vertex set.  A
{\bf $\mathbf{k}$-colouring} $\gamma$ of $D$ is a {\bf
  $\mathbf{k}$-out-colouring} if no out-neighbourhood is
monochromatic, that is for every vertex $u$ of $D$ there exist two
out-neighbours $v$ and $w$ of $u$ with $\gamma(u)\neq \gamma(v)$. For
any $k$-colouring of $D$ a vertex $x$ will be {\bf good} if its
out-neighbourhood is not monochromatic.  So a $k$-colouring is a
$k$-out-colouring if all vertices of $D$ are good.

For a 2-colouring $\gamma:V\rightarrow \{1,2\}$ of a digraph $D$ we
will denote by $V_i$ the set $\{x\in V\ : \ \gamma(x)=i\}$, for
$i=1,2$.  As the set $V_1$ totally defines $\gamma$ we will often just
specify it to describe the 2-colouring $\gamma$ and we will simply say
that it {\bf defines} $\gamma$.

The following observation shows that even for some classes of highly
structured digraphs, for every integer $r$ there does not exist a
number $K$ so that all digraphs from this class and with minimum
out-degree at least $K$ have an $r$-out-colouring. A {\bf bipartite
  tournament} is a digraph that can be obtained from a complete
bipartite graph $B$ by assigning an orientation to each edge of $B$.

\begin{proposition}
  \label{nodegreeboundorcolbound}
For all positive integers $k,r$ there exists a $k$-strong bipartite tournament
$B_{k,r}$ with $\delta^+(B_{k,r})=k$ which has no 
$r$-out-colouring.
\end{proposition}

\pf Let $U$ be an independent set on $kr$ vertices, let
$X_1,X_2,\ldots{},X_{kr\choose k}$ be an ordering of the distinct
$k$-subsets of $U$ and let $V$ be an independent set of $kr\choose k$
vertices $v_1,v_2,\ldots{},v_{kr\choose k}$ where $v_i$ corresponds to
$X_i$ for each $i\in [{kr\choose k}]$. The vertex set of $B_{k,r}$ is
$U\cup V$ and the arc set consist of the arcs from $v_i$ to $X_i$,
$i\in [{kr\choose k}]$ and all other arcs between vertices in $U$ and
$V$ go from $U$ to $V$. It is easy to check that $B_{k,r}$ is $k$-strong.
No matter how we $r$-colour $U$ there will be
a monochromatic $k$-set $X_i$, showing that $B_{k,r}$ has no
$r$-out-colouring. \qed

By adding a new set $W$ inducing a digraph with chromatic number at
least $p-2$ and all possible arcs from $U$ to $W$ and from $W$ to $V$,
we see that there is no bound on the out-degree that guarantees a
$r$-out-colouring, even when the digraph is $p$-chromatic.\\

We shall use the following classical result due to Moon.
\begin{theorem}[Moon]\cite{moonCMB9}
\label{moonthm}
Every strong tournament is vertex-pancyclic.
\end{theorem}

\section{Complexity of 2-out-colouring}\label{complexitysec}

Every undirected graph $G=(V,E)$ corresponds to the symmetric digraph
$D=\stackrel{\leftrightarrow}{G}$ that we obtain from $G$ by replacing
every edge of $G$ by a directed 2-cycle.

A {\bf total dominating} set of a graph $G=(V,E)$ is a set of vertices
$S$ such that every vertex of $G$ has a neighbour in $S$. Note that a
graph $G$ has a partition into two total dominating sets $(S,V-S)$ if
and only if $\stackrel{\leftrightarrow}{G}$ has a
2-out-colouring. Thus the result of \cite{heggernesNJC5} that
deciding whether a graph has a 2-partition into total dominating sets
is ${\cal NP}$-complete, implies the following.

\begin{theorem}
  \label{2outcolnpc}
Deciding whether a given digraph has a 2-out-colouring is ${\cal NP}$-complete.
\end{theorem}

For completeness we give a short proof of the following strengthening
of Theorem \ref{2outcolnpc}

\begin{theorem}
For every integer $K$, deciding whether a given symmetric digraph of
minimum out-degree at least $K$, has a 2-out-colouring is ${\cal
  NP}$-complete.
\end{theorem}

\pf Let ${\cal H}=(X,E)$ be a hypergraph. We consider its bipartite
representation $G$. That is, the vertices of $G$ are $X\cup E$ and
there is an edge from $x\in X$ to $e\in E$ iff $x\in e$.  We add to
$G$ a vertex $z$ linked to all vertices of $X$ and then replace every
edge by a 2-cycle to obtain the digraph $D_{\cal H}$. It is clear that
$D_{\cal H}$ is a symmetric digraph, and we claim that ${\cal H}$ is
2-colourable if and only if $D_{\cal H}$ admits a
2-out-colouring. Indeed if ${\cal H}$ is 2-colourable then we keep
this colouring on the set of vertices of $D_{\cal H}$ corresponding to
$X$, colour the vertices of $D_{\cal H}$ corresponding to $E$ by 1 and
$z$ by 2. Then we obtain a 2-out-colouring of $D_{\cal
  H}$. Conversely, if $D_{\cal H}$ has a 2-out-colouring, as the
out-neighbours of every vertex belonging to $E$ contains different
colours, the colouring of $X$ corresponds to a 2-colouring of the
hypergraph $\cal H$.\\ The claim now follows from the ${\cal
  NP}$-completeness of the 2-coloring problem for
hypergraphs~\cite{lovasz1973}, mentioned in the introduction, and the
easy fact that the hypergraph 2-colouring problem remains ${\cal
  NP}$-complete when every vertex is in at least $K$ hyperedges and
all hyperedges have size at least $K$.  \qed

We show below that even for highly structured digraphs with many arcs,
the problem is still hard.

\begin{theorem}
  It is NP-complete to decide whether a bipartite tournament with
  minimum out-degree 3 admits a 2-out-colouring.
\end{theorem}

\pf Let $\cal F$ be an instance of {\sc monotone
  NAE-3-SAT}\footnote{Recall that NAE-3-SAT is the variant of
    3-SAT where we seek a truth assignment $t$ such that each clause
    has both a false and a true literal under $t$. The further
    restriction monotone NAE-3-SAT means that we restrict to instances
    with no negated variables. This problem is still ${\cal
      NP}$-complete \cite{shaeferSTOC10}.} with variables
$x_1,x_2,\ldots{},x_n$ and clauses $C_1,C_2,\ldots{},C_m$. By
reordering if necessary, we may assume that $C_1$ and $C_2$ contain no
common variable.  Let $U=\{u_1,u_2,\ldots{}, u_n\}$ and
$V=\{c_1,c^*_1,c^{**}_1,c_2,c^*_2,c^{**}_2,c_3,\ldots{},c_{m-1},c_m,\}$
be two disjoint vertex sets. For each clause $C_i$, $i\in [m]$ we add
3 arcs from $c_i$ to the vertices $u_{i_1},u_{i_2},u_{i_3}$
corresponding to the three literals $x_{i_1},x_{i_2},x_{i_3}$ of $C_i$
and let all remaining vertices in $U$ have an arc to $c_i$. Finally we
let the arcs incident to $c^*_i,c^{**}_i$ be copies of the arcs
incident to $c_i$ for $i=1,2$ (so we have 3 vertices for each of
$C_1,C_2$ and one for each other clause).\\ We claim that the
resulting bipartite tournament $B=B({\cal F})$ has a 2-out-colouring
if and only there is a truth assignment $\phi$ to the variables such
that every clause has either one or two true literals. Suppose first
that $B$ has a 2-out-colouring. Then we let variable $x_i$ be true
precisely if $u_i$ receives colour 1 under this colouring. Since $c_j$
has precisely 3 out-neighbours in $B$ it follows that this truth
assignment will satisfy one or two variables of each
clause. Conversely, suppose $\phi$ is a truth assignment to the
variables so that $C_j$ has either one or two true literals for $j\in
[m]$. For each $i\in [n]$ such that $\phi{}(x_i)=true$ we colour $u_i$
by 1 and for all $i$ such that $\phi{}(x_i)=false$ we colour $u_i$ by
2. Now all clause vertices have both colours in their
out-neighbourhoods, so we just have to make sure the same holds for
the vertices $u_i$, $i\in [n]$. We obtain this by colouring all $c_j$,
$j\in [m]$ as well as $c^*_1,c^*_2$ by colour 1 and
$c^{**}_1,c^{**}_2$ by colour 2. Since every $u_i$ either dominates
both of $c^*_1,c^{**}_1$ or both of $c^*_2,c^{**}_2$, every vertex of
$U$ is good also and this colouring is a 2-out-colouring of $B({\cal
  F})$. \qed

\section{Out-colourings of tournaments}\label{Toutcolsec}

In this section we focus on 2-out-colourings of tournaments.

Let $T$ be a tournament which is not strongly connected and denote by
$C$ the terminal component of $T$. As $T\setminus C\Rightarrow C$ the
following holds.

\begin{observation}
\label{observation:strong}
If $\gamma$ is a $k$-out-colouring of the terminal component $C$ of a
tournament $T$, then every extension of $\gamma$ to the vertices of
$T\setminus C$ leads to a $k$-out-colouring of $T$.
\end{observation}

\begin{lemma}
\label{lemma:indomnating-cycle}
Every tournament $T$ on $n\ge 5$ vertices contains either an
in-dominating vertex or an in-dominating cycle of size at most $n-2$.
\end{lemma}

\pf First assume that $T$ is not strongly connected and denote by $X$
the terminal component of $X$. If $X$ only contains one vertex, then
this vertex is an in-dominating vertex of $T$. Otherwise if $|X|\ge 4$
then by Theorem \ref{moonthm}, $T$ contains a cycle on $|X|-1$
vertices which is an in-dominating cycle of $T$ of size at most
$n-2$. And if $|X|=3$ then a Hamiltonian 3-cycle of $X$ forms also an
in-dominating cycle of $T$ of size at most $n-2$.\\ Now if $T$ is
strongly connected, by Theorem~\ref{moonthm} it contains a
cycle $C$ on $n-2$ vertices. Let us denote by $x$ and $y$ the vertices
of $T\setminus C$.  If $x$ and $y$ have both an out-neighbour on $C$,
then $C$ is an in-dominating cycle of $T$. Otherwise, as $T$ is
strongly connected, it means for instance that $y$ has an
out-neighbour on $C$ and that we have $x\dom y$ and $C\sdom x$. In
this case if we let $z$ be an out-neighbour of $y$ on $C$, $xyz$ is an
in-dominating cycle of $T$ of size $3\le n-2$.  \qed

\subsection{2-out-colouring of tournaments with minimum out-degree 2}

The {\bf rotational tournament on 5 vertices} denoted by $RT_5$ has
vertex set $\{1,2,3,4,5\}$ and is the union of the two directed cycles
$12345$ and $13524$. It is easy to check that $RT_5$ has no
2-out-colouring. Moreover we define the tournament $T_7$ on 7
vertices. It contains two 3-cycles $C$ and $C'$ and a vertex $z$ such
that $C'\Rightarrow z$, $z\Rightarrow C$ and $C\Rightarrow C'$. It is
also easy to check that $T_7$ has no 2-out-colouring.\\ A tournament
$T$ on $n\ge 6$ vertices with $\delta^+(T)= 2$ belongs to {\bf the
  family ${\cal G}_1$} if there exists a sub-tournament $T'$ of $T$ on
$n-3$ vertices such that: $T'$ has an in-dominating vertex $w$,
$T\setminus T'$ is a 3-cycle $C$, $C\Rightarrow z$ where $z$ is a
vertex of $T'$ different from $w$ and with out-degree at least 2, and
$T'\setminus z \Rightarrow C$.  Notice that if $T$ belongs to ${\cal
  G}_1$ then $T$ has no 2-out-colouring.  Indeed in such a colouring
all the vertices of $C$ must receive the same colour (different from
the colour of $z$) and then the out-neighbourhood of $w$ would be
monochromatic. The tournaments $RT_5$, $T_7$ and the family ${\cal
  G}_1$ are depicted in Figure~\ref{fig:semicompdelat2} (for the
more general case of semicomplete digraphs).

\begin{theorem}
\label{theo:delta+=2}
A tournament $T$ with $\delta^+(T)= 2$ admits a 2-out-colouring except
if it belongs to the family ${\cal G}_1$ or if its terminal strong
component is $RT_5$ or $T_7$.
\end{theorem}

\pf By Observation~\ref{observation:strong} we can assume that $T$ is
strongly connected. We deal with two different cases.

{\it Case 1: $T$ has exactly one vertex with out-degree 2}. In this
case $T$ admits a 2-out-colouring.  Indeed denote by $a$ the vertex of
$T$ with out-degree 2 and by $b$ and $c$ its out-neighbours with
$b\rightarrow c$. Let also $c_1$ and $c_2$ be two out-neighbours of
$c$. Consider the 2-colouring defined by $\{a,c,c_1\}$.  As every
vertex different from $a$ has degree at least 3, it is a
2-out-colouring of $T$ except if there exists a vertex $d_1$
whose out-neighbourhood is exactly $\{a,c,c_1\}$. Now the 2-colouring
defined by $\{a,c,c_2\}$ is a 2-out-colouring of $T$: If this was not
the case, then there would be a new vertex $d_2$ with
$N^+(d_2)=\{a,c,c_2\}$, but then $d_1$ dominates $d_2$, contradicting
our conclusion above.

{\it Case 2: $T$ contains at least two vertices of degree 2}. We
observe two different sub-cases.\\ {\it Case 2.1: No two vertices of
  out-degree 2 have a common out-neighbour}. In this case, denote by
$a$ and $b$ two vertices of out-degree 2 with $a\rightarrow b$. Denote
by $c$ the second out-neighbour of $a$ and by $d$ and $e$ the two
out-neighbours of $b$ with $d\rightarrow e$. Notice that we have
$c\rightarrow b$ and $\{d,e\}\Rightarrow a$. First consider the
2-colouring of $T$ defined by $\{a,b,e\}$. It is a 2-out-colouring of
$T$ except if there exists a vertex $x$ with its out-neighbours
included in $\{a,b,e\}$.  In this case the out-neighbourhood of $x$ is
exactly $\{a,b,e\}$ otherwise $x$ has out-degree 2 and has a common
out-neighbour with $a$ or $b$ which is excluded in this sub-case. In
particular we have $x\in N^-(a)\cup N^-(b)$ and every vertex different
from $a$, $b$ and $e$ dominates $x$. So now the 2-colouring with first
set $\{a,b,d\}$ is a 2-out-colouring of $T$.\\ {\it Case 2.2: There
  exist two vertices of out-degree 2 with a common out-neighbour}.
Call by $a$ and $b$ two such vertices with $a\rightarrow b$ and denote
by $d$ their common out-neighbour and by $c$ the second out-neighbour
of $b$. Also denote by $X$ the set $N^-(a)\cap N^-(b)=V(T)\setminus
\{a,b,c,d\}$. First assume that $c\rightarrow d$.  If $c$ has an
out-neighbour $e$ in $X$ and $d\rightarrow e$ then $\{b,c,e\}$ defines
a 2-out-colouring of $T$. If $d$ does not dominate $e$ then
$\{b,c,e,f\}$ defines a 2-out-colouring of $T$ where $f$ is an
out-neighbour of $d$. So we can assume that $X\Rightarrow \{a,b,c\}$
and that $c$ has out-degree 2 also. If there is $w\in X$ whose
out-neighbourhood is exactly $\{a,b,c\}$ then $T$ belongs to the
family ${\cal G}_1$, with $T'=T[X\cup \{d\}]$, $C=abc$ and $d$ playing
the role of $z$. Otherwise let $e_3$ be an out-neighbourhood of $d$
and consider the 2-colouring defined by $\{a,b,c, e_3\}$. If it is not
a 2-out-colouring of $T$, it means that there exist $e_2$ whose
out-neighbourhood is exactly $\{a,b,c,e_3\}$ (it cannot be just
$\{a,b,c\}$ otherwise $T$ belongs to ${\cal G}_1$). Similarly if
$\{a,b,c,e_2\}$ does not define a 2-out-colouring of $T$ there exists
$e_1$ whose out-neighbourhood is $\{a,b,c,e_2\}$. Now either
$\{a,b,c,e_1\}$ defines a 2-out-colouring of $T$ or
$T[a,b,c,d,e_1,e_2,e_3]$ is the terminal component of $T$ and
isomorphic to $T_7$. In this later case $T$ does not admit a
2-out-colouring.\\ Finally, we treat the case where $d\rightarrow
c$. Assume that there exist $e$ and $f$ two different vertices in $X$
such that $c\rightarrow e$ and $d\rightarrow f$. Then $\{b,c,e\}$
defines a 2-out-colouring of $T$.  Thus we may assume that $c$
and $d$ are vertices of out-degree 2 and they have a common
out-neighbour $e$ in $X$. If $X\setminus \{e\}=\emptyset$ then
$T=RT_5$, otherwise $X\setminus \{e\}\Rightarrow \{a,b,c,d\}$ and as
$T$ is strongly connected $e$ must have an out-neighbour in $X$. So we
check that $\{a,b,c\}$ defines a 2-out-colouring of $T$.  \qed

\subsection{2-out-colouring of tournaments with minimum out-degree at least 3}

\begin{lemma}
\label{lemma:dominant-of-size-2}
Every tournament $T$ with $\delta^+(T)\ge 3$ and which has an
in-dominating set of size 2 admits a 2-out-colouring
\end{lemma}

\pf Assume that $\{a,b\}$ is an in-dominating set of $T$ with $a$
dominating $b$.  Let $c$ be an out-neighbour of $a$.  We consider the
2-colouring of $T$ defined by $\{a,b,c\}$. If it is not a
2-out-colouring of $T$ it means that there exists a vertex $d$ whose
out-neighbourhood is exactly $\{a,b,c\}$. So we choose another
out-neighbour $c'$ of $b$ and check that $\{a,b,c'\}$ defines a
2-out-colouring of $T$ (as every vertex of $T\setminus \{a,b,c,d\}$
dominates either $a$ or $b$ and dominates $d$).  \qed

The {\bf Paley tournament on 7 vertices} denoted by $P_7$ has vertex
set $\{1,2,3,4,5,6,7\}$ and is the union of the three directed cycles
$1234567$, $1357246$ and $1526374$. It is easy to check that $P_7$ has
no 2-out-colouring.

 \begin{theorem}
\label{theo:mindegree3}
  Every tournament $T$ with $\delta^+(T)\ge 3$ and whose terminal
  strong component is different from $P_7$ admits a 2-out-colouring.
\end{theorem}

\pf By Observation~\ref{observation:strong} we can assume that $T$ is
strongly connected. Consider a vertex $x$ of $T$ with minimum
out-degree. If $T[N^+(x)]$ has an in-dominating vertex $y$, then
$\{x,y\}$ is an in-dominating set of $T$ of size 2 and we conclude
with Lemma~\ref{lemma:dominant-of-size-2}. We study now different
cases:

First assume that $\delta^+(T)\ge 5$. In particular $d^+(x)\ge 5$ and
by Lemma~\ref{lemma:indomnating-cycle} $T[N^+(x)]$ contains an
in-dominating cycle $C$ of size at most $d^+(x)-2$. So $C\cup x$
defines a 2-out-colouring of $T$. Indeed it is clear that every vertex
has an out-neighbour coloured by 1. And if a vertex $y$ has only
out-neighbours coloured by 1, it means that $d^+(y)\le |C|+1\le
d^+(x)-2+1<d^+(x)$, a contradiction.

Now assume that $\delta^+(T)=4$. We have $d^+(x)=4$ and as $T[N^+(x)]$
does not contain any in-dominating vertex, $T[N^+(x)]$ contains an
in-dominating 3-cycle. So we denote $N^+(x)=\{a,b,c,d\}$ with $abc$
being a 3-cycle and $d\rightarrow a$. We consider the 2-colouring of
$T$ defined by $\{x,a,b,c\}$. If it is not a 2-out-colouring it means
that there exists $y\in N^-(x)$ such that $N^+(y)=\{x,a,b,c\}$. In
this case let $e$ be an out-neighbour of $b$ lying in $N^-(x)$ and
consider the 2-colouring of $T$ defined by $\{x,a,b,e\}$. As every
vertex of $N^-(x)\setminus y$ dominates $y$, this colouring is a
2-out-colouring of $T$.

Finally we assume that $\delta^+(T)=3$. We have $d^+(x)=3$ and as
$T[N^+(x)]$ does not contain any in-dominating vertex, $T[N^+(x)]$ is
3-cycle, denoted by $abc$. By Lemma \ref{lemma:indomnating-cycle}, we
may assume that $N^+(a)\cap N^+(b)\neq\emptyset$, $N^+(b)\cap
N^+(c)\neq\emptyset$ $N^+(c)\cap N^+(a)\neq\emptyset$. If $a$, $b$ and
$c$ have a common out-neighbour $y$ in $N^-(x)$ then consider the
2-out-colouring of $T$ defined by $\{x,a,y\}$. If it is not a
2-out-colouring of $T$, then $\{x,b,y\}$ defined a 2-out-colouring of
$T$.  Similarly if $a$ and $b$ have two common out-neighbours $y$ and
$y'$, then $\{x,a,y\}$ or $\{x,a,y'\}$ defines a 2-out-colouring of
$T$.  Thus by symmetry, it means that $a$ and $b$ have exactly one
common out-neighbour, called $d$, that $b$ and $c$ have exactly one
common out-neighbour, called $f$ and that $c$ and $a$ have exactly one
common out-neighbour, called $e$.  Using that every vertex of
$N^-(x)\setminus d$ dominates $a$ or $b$, we can deduce the following.
If $d\rightarrow f$ then $\{x,c,d,e\}$ defines a 2-out-colouring of
$T$. So by symmetry we can assume that $def$ is a 3-cycle of $T$. If
$\{a,b,c\}$ has an out-neighbour in $N^-(x)\setminus \{d,e,f\}$, say
that $c\rightarrow y$ with $y\in N^-(x)\setminus \{d,e,f\}$, then
$\{x,c,d,y\}$ defines a 2-out-colouring of $T$. Finally if
$N^-(x)\setminus \{d,e,f\}\neq \emptyset$ then as $T$ is strongly
connected it means that for instance $d\rightarrow y$ for some $y\in
N^-(x)\setminus \{d,e,f\}$. But then $\{x,c,d,e\}$ defines a
2-out-colouring of $T$. Otherwise $N^-(x)\setminus \{d,e,f\}=
\emptyset$ and $T$ is exactly the Paley tournament.  \qed

Together Theorem~\ref{theo:delta+=2}, Theorem~\ref{theo:mindegree3}
and the fact that our proofs are constructive provide the following
corollary.

\begin{corollary}
  A tournament $T$ with $\delta^+(T)\ge 2$ admits a 2-out-colouring if
  and only if its terminal strong component is different from $RT_5$,
  $T_7$ and $P_7$ and $T$ does not belong to the family ${\cal
    G}_1$. Consequently, in polynomial time, one can decide whether or
  not a tournament admits a 2-out-colouring and find a 2-out-colouring
  when it exists.
\end{corollary}

\section{Out-colourings of semicomplete digraphs}\label{SDoutcolsec}

We extend the results of the previous section to semicomplete digraphs.

\subsection{2-out-colourings of semicomplete digraphs with minimum out-degree 2}

\begin{figure}[!ht]
\centering
\scalebox{0.5}{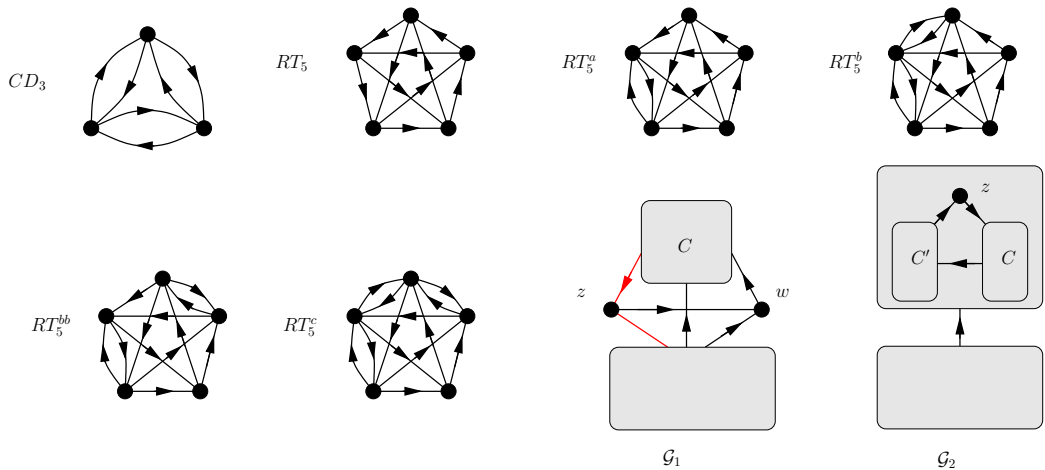}
\caption{The semicomplete digraphs $CD_3$, $RT_5$, $RT_5^a$, $RT_5^b$,
  $RT_5^{bb}$ and $RT_5^c$ and the families ${\cal G}_1$ and ${\cal
    G}_2$ (red arcs could be replaced by 2-cycles and non oriented red
  edges could be oriented in any direction).}
\label{fig:semicompdelat2}
\end{figure}

Figure~\ref{fig:semicompdelat2} defines a family $\cal G$ of
semicomplete digraphs. A digraph belongs to $\cal G$ if its terminal
strong component is $CD3$, $RT_5$, $RT_5^a$, $RT_5^b$, $RT_5^{bb}$ or
$RT_5^c$ or if it belongs to ${\cal G}_1$ or ${\cal G}_2$.  A digraph
belongs to ${\cal G}_1$ if it contains a vertex $w$ whose
out-neighbourhood is a 2- or 3-cycle $C$ such that the
out-neighbourhood of $V(C)$ is exactly one vertex $z$ different from
$w$.  Finally a semicomplete digraph $T$ belongs to the family ${\cal
  G}_2$ if its terminal strong component contains a 2- or 3-cycle $C$
which dominates a 2- or 3-cycle $C'$ which dominates a vertex $z$
dominating $C$.

\begin{lemma}
\label{lemma:semicompdelat2}
None of the semicomplete digraphs in $\cal G$ admit a
2-out-colouring. Moreover for each semicomplete digraph $T\in{\cal G}$
if we add an arc to $T$, then either the resulting digraph admits a
2-out-colouring or it belongs to $\cal G$.
\end{lemma}

\pf We will not provide the complete proof of the statement but just
give some important remarks on how to obtain the result.  We leave it
to the reader to check that if we add an extra arc to one of the six
digraphs in ${\cal G}-{\cal G}_1-{\cal G}_2$, then we either get a new
digraph in ${\cal G}$ or the resulting digraph has a 2-out-colouring.
However we pay attention to the family ${\cal G}_1$ where the most
technical case occurs. Let $T$ be a digraph of ${\cal G}_1$ and $xy$
be an arc to add to $T$. If $xy$ is incident neither to $C$ nor to $w$
then $T+xy$ is clearly a member of ${\cal G}_1$. The same holds if
$x=z$ and $y\in C$. If $x\in C$ or $xy=wz$ then we check that $T+xy$
admits a 2-out-colouring. Finally let us assume that $x=w$ and that
$y$ belongs to $X=V(T)\setminus (V(C)\cup \{w,z\})$. If $z$ has a
second out-neighbour in $X$ different from $y$ then $\{w,y,z\}$
defines a 2-out-colouring of $T+xy$. So assume now that $y$ is the
only out-neighbour of $z$ in $X$. If $y$ has an out-neighbour $t$ in
$X$ then $\{z,y,t\}$ defines a 2-out-colouring of $T+xy$. In the
remaining case $V(C)\cup \{w,y,z\}$ forms the terminal strong
component of $T+xy$ and contains the 2-cycle $wy$ which dominates $C$
which dominates $z$ which in turn dominates the 2-cycle $wy$. Then
$T+xy$ belongs to ${\cal G}_2$. Finally let us just mention that
adding an arc to any digraph of ${\cal G}_2$ leads to a digraph
admitting a 2-out-colouring.  \qed

\begin{theorem}
\label{theo:SemicompDelta2}
A semicomplete digraph $T$ with $\delta^+(T)=2$ admits a
2-out-colouring unless its terminal strong component is $CD_3$, $RT_5$,
$RT_5^a$, $RT_5^b$, $RT_5^{bb}$ or $RT_5^c$ or if it belongs to ${\cal
  G}_1$ or ${\cal G}_2$.
\end{theorem}

\pf We prove the result by induction on the number of 2-cycles in
$T$. If $T$ does not contain any 2-cycle then the result holds by
Theorem~\ref{theo:delta+=2}. Assume now that $T$ has a 2-cycle $xy$
and suppose first that the digraph $T-xy$ still satisfies
$\delta^+(T-xy)=2$. If $T-xy$ admits a 2-out-colouring then so does
$T$. Otherwise by induction it means that $T-xy$ belongs to the family
$\cal G$. By Lemma~\ref{lemma:semicompdelat2} $T$ also belongs to
$\cal G$ or admits a 2-out-colouring.

Now by symmetry we can assume that both $x$ and $y$ have out-degree
exactly 2 in $T$. We look at two sub-cases: either $x$ and $y$ have
common out-neighbour or not. Assume first that $x$ and $y$ both
dominate a vertex $z$. If there exists $w\in V(T)$ such that
$N^+_T(w)=\{x,y\}$, then either $w=z$ and the terminal component of
$T$ is $CD_3$ or $w\neq z$ and $T$ belongs to the family ${\cal
  G}_1$. In both cases $T$ does not admit a 2-out-colouring. Otherwise
it means that $\delta^+(T\setminus \{x,y\})\ge 1$. If $z$ dominates
$x$ or $y$ then $\{x,y\}$ defines a 2-out-colouring of $T$. If $z$ has
an out-neighbour $u$ such that $\delta^+(T\setminus \{x,y,u\})\ge 1$
then $\{x,y,u\}$ defines a 2-out-colouring of $T$. Otherwise it is
easy to check that the out-neighbourhood of $z$ is exactly a 2-
or a 3-cycle $C$ dominated by $V(T)\setminus (V(C)\cup
\{x,y\})$. But in this case $T$ belongs to ${\cal G}_2$.\\ For the
other sub-case, assume that $x$ and $y$ do not have a common
out-neighbour.  So we respectively denote by $z$ and $t$ the second
out-neighbour of $x$ and $y$. In particular, $z$ dominates $y$ and $t$
dominates $x$. Without loss of generality we can assume that $z$
dominates $t$. If $\delta^+(T\setminus \{x,y\})\ge 1$, then $\{x,y\}$
defines a 2-out-colouring of $T$. Otherwise it means that there exists
$v\in T$ with $N^+(v)=\{x,y\}$. Notice that $v\neq z$ as $z\dom t$. We
treat together the cases $v=t$ and $v\neq t$.  If $z$ has an
out-neighbour $u$ in $V(T)\setminus \{x,y,t,v\}$ then $\{x,z,u\}$
defines a 2-out-colouring of $T$. Otherwise the out-neighbourhood of
$z$ is exactly $\{v,t,y\}$ and $T$ belongs to the family ${\cal G}_1$
with $tvy$ being the 2- or 3-cycle $C$ and $x$ playing the role of
$w$. \qed

\subsection{2-out-colourings of semicomplete digraphs with minimum out-degree at least 3}

\begin{lemma}
\label{lemma:indomSemiComp}
  Every semicomplete digraph $T$ with $\delta^+(T)\ge 3$ which admits
  an in-dominating set of size 2 has a 2-out-colouring.
\end{lemma}

\pf Let $\{a,b\}$ be an in-dominating set of $T$. If $T[a,b]$ is a
2-cycle, then $\{a,b\}$ defines a 2-out-colouring of $T$, as
$\delta^+(T)\ge 3$. Otherwise assume that $a$ dominates $b$ and that
$b$ does not dominate $a$. Let $c$ be an out-neighbour of $b$. We know
that $c$ dominates $a$ or $b$ and so if $\{a,b,c\}$ does not define a
2-out-colouring of $T$, it means that there exists $d\in V(T)$ such
that $N^+(d)=\{a,b,c\}$. As $a$ is not an out-neighbour of $b$, there
exists a vertex $c'\notin \{a,b,c,d\}$ such that $b \dom c'$.  As
previously if $\{a,b,c'\}$ does not define a 2-out-colouring of $T$
there exists $d'\in V(T)$ with $N^+(d')=\{a,b,c'\}$. As $d'$ does not
dominate $d$ we must have $d'=c$. Now if $b\dom d$ then $bd$ is a
in-dominating 2-cycle of $T$, and we conclude as in the first part of
the proof. Otherwise there exists $c''\notin \{a,b,c,d,c'\}$ such that
$b\dom c''$. To conclude we check that $\{a,b,c''\}$ defines a
2-out-colouring of $T$.\qed

\begin{theorem}
\label{theo:SemicompDelta3}
  Every semicomplete digraph $T$ with $\delta^+(T)\ge 3$ admits a
  2-out-colouring unless its terminal strong component is $P_7$.
\end{theorem}

\pf We prove the result by induction on the number of 2-cycles in
$T$. If $T$ does not contain any 2-cycle then the result holds by
Theorem~\ref{theo:mindegree3}. Besides it is easy to show the Paley
tournament on 7 vertices plus one arc admits a 2-out-colouring. Indeed
without loss of generality, this semicomplete digraph has vertex set
$\{1,\dots ,7\}$ and arc set $\{ij\ : \ j-i=1, 2\textrm{ or } 4 \mod
7\}\cup \{21\}$ and admits the 2-out-colouring defined by
$\{3,4,6\}$. So assume that $T$ contains a 2-cycle $ab$. When we
remove from $T$ the arc $ab$ if we obtain a semicomplete digraph with
minimum out-degree at least 3 (including $P_7$) then $T$ admits a
2-out-colouring by induction or by the previous remark. Otherwise it
means that $a$ has out-degree exactly 3. The same holds for $b$ and we
denote by $S$ the set $(N^+(a)\cup N^+(b))\setminus \{a,b\}$. Thus we
have $2\le |S|\le 4$. If we have $|S|=4$ then $S$ is an in-dominating
set of $T$ of size 2 and we conclude with
Lemma~\ref{lemma:indomSemiComp}. If $|S|=2$ then $a$ and $b$ have two
common out-neighbours $c$ and $d$. Assume that $c$ dominates $d$, then
$\{a,d\}$ is an in-dominating set of $T$ of size 2 and we conclude
again with Lemma~\ref{lemma:indomSemiComp}. Finally if $|S|=3$ then
$a$ and $b$ has a common out-neighbour $c$ and we denote by $d$ the
third out-neighbour of $a$ and by $e$ the one of $b$. As $|S|=3$ we
have $d\neq e$. If $d$ and $e$ both dominate $c$, then $\{a,c\}$ is an
in-dominating set of $T$ an we use
Lemma~\ref{lemma:indomSemiComp}. Otherwise we can assume without loss
of generality that $c$ dominates $d$. As $a$ does not dominate $e$,
$e$ has to dominate $a$. Once again $\{a,d\}$ is an in-dominating set
of $T$ and we use Lemma~\ref{lemma:indomSemiComp} to conclude.  \qed

As the proofs of Theorem~\ref{theo:SemicompDelta2} and
Theorem~\ref{theo:SemicompDelta3} are constructive provide we obtain
the following corollary.

\begin{corollary}
A semicomplete digraph $T$ with $\delta^+(T)\ge 2$ admits a
2-out-colouring unless its terminal strong component is $CD_3$,
$RT_5$, $RT_5^a$, $RT_5^b$, $RT_5^{bb}$, $RT_5^c$ or $P_7$ or if it
belongs to ${\cal G}_1$ or ${\cal G}_2$. Consequently, in polynomial
time, one can decide whether or not a semicomplete digraph admits a
2-out-colouring and find a 2-out-colouring when it exists.
\end{corollary}

To conclude this section, notice that every exception listed in the
previous statement can be 3-out-coloured. So we obtain the following.

\begin{corollary}
Every semicomplete digraph $T$ with $\delta^+(T)\ge 2$ admits a
3-out-colouring.
\end{corollary}

\section{Balanced 2-out-colourings of semicomplete digraphs}\label{Balanced}

A 2-out-colouring with colour classes $V_1$ and $V_2$ is {\bf
  balanced} if we have $\bigl| |V_1|-|V_2|\bigr| \le 1$.\\ The next
figure defines three families of semicomplete digraphs ${\cal T}_1$,
${\cal T}_2$ and ${\cal T}_3$ containing respectively two, three and
five digraphs. It is not difficult to check that all of the digraphs
in ${\cal T}_1$, ${\cal T}_2$ and ${\cal T}_3$ have a 2-out-colouring
(the one indicated by the two columns of vertices) and that none of
them have a balanced 2-out-colouring.
\begin{figure}[!ht]
\centering
\scalebox{0.6}{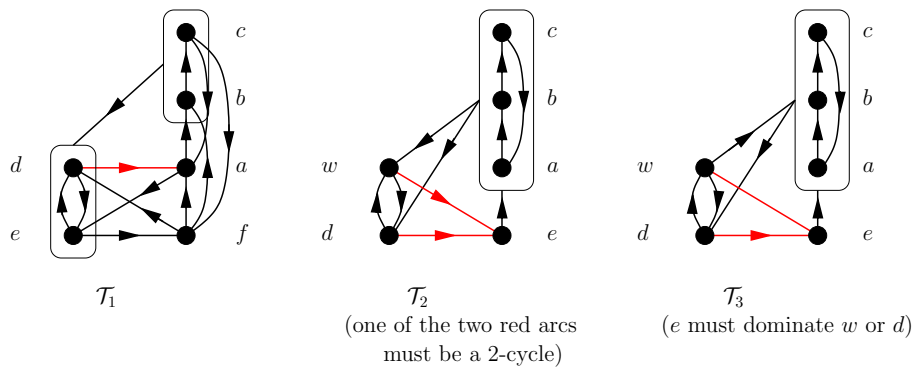}
\caption{The families ${\cal T}_1$, ${\cal T}_2$ and ${\cal T}_3$ of
  semicomplete digraphs which admit a 2-out-colouring but no balanced
  2-out-colouring (red arcs could be replaced by 2-cycles and non
  oriented red edge could be oriented in any direction).}
\label{fig:Balanced}
\end{figure}

\begin{theorem}
\label{theo:balanced}
Every semicomplete digraph $T$ which admits a 2-out-colouring also
admits a balanced 2-out-colouring except if $T$ belongs to ${\cal
  T}_1$, ${\cal T}_2$ or ${\cal T}_3$.
\end{theorem}

\pf Let $T$ be a semicomplete digraph which admits a 
2-out-colouring. If $|T|\le 5$ then any 2-out-colouring of $T$ is
balanced. So we assume that $|T|\ge 6$ and among all the
2-out-colourings of $T$ we consider one, denoted by $\gamma$, inducing a
partition $(V_1,V_2)$ with $\bigl| |V_1|-|V_2|\bigr|$ as small as
possible. Assume that $\bigl| |V_1|-|V_2|\bigr|>1$ and that we have
$|V_2|> |V_1|+1$. So, let $Y_2$ be the set $\{v_2\in V_2\ :
\ d^+_{V_2}(v_2)=1\}$, let $X_2$ be $N^+_{V_2}(Y_2)$ and for every
vertex $v_2\in V_2$ let $Z_{v_2}$ be the set $\{v_1\in
V_1\ :\ N^+_{V_2}(v_1)=\{v_2\}\}$. Notice that as $T$ is
semicomplete, we have $|X_2|=|Y_2|\le 3$ and that if a vertex $v_2\in
V_2\setminus X_2$ satisfies $Z_{v_2}=\emptyset$ then $V_1\cup\{v_2\}$
also defines a 2-out-colouring of $T$, contradicting the choice of
$\gamma$. So we have $Z_{v_2}\neq \emptyset$ for every $v_2\in V_2\setminus
X_2$. Moreover, by definition, all the sets $Z_{v_2}$ are disjoint and we
have $|V_1|\ge \sum_{v_2\in V_2\setminus X_2}|Z_{v_2}|\ge |V_2\setminus
X_2|=|V_2|-|X_2|$. So we have $|X_2|=2$ or $|X_2|=3$.

\SB{Below we will show that we can either easily reach a contradiction to
the optimality of $\gamma$ or we may choose one vertex $x$ in $V_1$
and two vertices $y$ and $z$ in $V_2$ such that the 2-colouring
$\gamma{}'$ defined by $(V_1\setminus \{x\})\cup \{y,z\}$ is a
2-out-colouring of $T$, contradicting also the choice of $\gamma$.} In
order to define $\gamma{}'$ we will also select another vertex $x'$ in
$V_1$. If $V_1$ contains a vertex with out-degree exactly 1 in $V_1$
then we choose $x'$ to be such a vertex and $x$ the only
out-neighbour of $x'$ in $V_1$. Otherwise we choose $x'$ to be any
vertex of $V_1$ and $x$ one out-neighbour of $x'$ in $V_1$. In any
case, every vertex of $V_1\setminus \{x'\}$ has at least one
out-neighbour in $V_1\setminus \{x\}$. Then we consider several cases.

First assume that $|Y_2|=2$ and that $T[Y_2]$ is 2-cycle $ab$.  In
this case we have $X_2=Y_2$ and $V_2\setminus X_2$ dominates
$X_2$. Thus we also have $|V_2|=|V_1|+2$ and $|Z_{v_2}|=1$ for every
vertex $v_2\in V_2\setminus X_2$ and then $V_1=\cup_{v_2\in
  V_2\setminus X_2} Z_{v_2}$. Notice that the arcs from $V_1$ to $V_2$
form a matching. So we choose $y\in V_2$ such that $Z_y=\{x'\}$ and
$z=a$. Let us check that in this case $\gamma{}'$ is 2-out-colouring
of $T$. Every vertex of $V_2\setminus X_2$ dominates $X_2$ and hence
dominates a vertex from $V_1(\gamma{}')$ and one from
$V_2(\gamma{}')$. The same holds for the vertices of $X_2$ because
they dominate $x'$ and $x$. Every vertex of $V_1\setminus x'$ has
still one out-neighbour in $V_1\setminus \{x\}\subset V_1(\gamma{}')$
and one in $V_2-\{y\}\subset V_2(\gamma{}')$. And finally $x'$
dominates $x$ which lies in $V_2(\gamma{}')$ and $y$ which lies in
$V_1(\gamma{}')$. So $\gamma{}'$ is a 2-out-colouring of $T$, a
contradiction.

Next assume that $|Y_2|=2$ also but that $T[Y_2]$ is not a 2-cycle.
In this case we must also have $|V_2|=|V_1|+2$ and $|Z_{v_2}|=1$ for
every vertex $v_2\in V_2\setminus X_2$ and then $V_1=\cup_{v_2\in
  V_2\setminus X_2} Z_{v_2}$. We denote by $a$ and $b$ the vertices of
$Y_2$ with $a\rightarrow b$ and we denote by $c$ the only
out-neighbour of $b$, that is, $X_2=\{b,c\}$. Notice that every vertex
of $V_2\setminus \{a,b,c\}$ dominates $\{a,b\}$.  If $|V_1|\ge 3$ then
every vertex of $V_2$ has at least two out-neighbours in $V_1$ and we
can choose $y\in V_2$ such that $Z_y=\{x'\}$ and $z=a$. As previously,
we check that $\gamma{}'$ is 2 out-colouring of $T$, using especially
that the only arcs from $V_1$ to $V_2$ is the matching
$\{uv\ :\ Z_v=\{u\}, v\in V_2\setminus X_2\}$ and so that every vertex
of $V_2$ has at least one out-neighbour in $V_1(c')$. So we must have
$|V_1|=2$ and then $V_1$ is a 2-cycle $de$ with $d\rightarrow a$ for
instance. Denote by $f$ the vertex of $V_2$ with $Z_{f}=\{e\}$. So $T$
must contain the following arcs: $fa$, $fb$, $ca$ and all the arcs
from $V_2$ to $V_1$ except possibly $ad$ and $fe$. If $T$ has a
2-out-colouring such that $d$ and $e$ receive the same colour,
then $a$, $b$, $c$ and $f$ are forced to receive the other colour and
the 2-out-colouring is not balanced. If $d$ and $e$ have different
colours in a balanced 2-out-colouring, then this 2-out-colouring must
have bipartition $(\{a,b,d\},\{c,e,f\})$.  Every vertex dominates a
vertex in each part except possibly $f$. But if $f$ does not dominate
$c$ or $e$ then $T$ belongs to the family ${\cal T}_1$.


Finally, assume that $|Y_2|=3$. In this case $Y_2$ is a 3-cycle,
denoted by $abc$, and $Y_2=X_2$ is dominated by $V_2\setminus X_2$. If
$|V_2\setminus X_2|=|V_1|$, then as previously, we have
$V_1=\cup_{v_2\in V_2\setminus X_2} Z_{v_2}$. We choose $x$, $x'$ and
$y$ as before and let $z=a$, then we check that the
  resulting colouring $\gamma{}'$ is a balanced 2-out-colouring of
$T$. Thus we must have $|V_1|=|V_2\setminus X_2|+1$.  In this
case we have $|Z_{v_2}|=1$ for every vertex $v_2\in V_2$ except
possibly for one vertex $v_2$ where we can have
$|Z_{v_2}|=2$. So we pick one vertex in each $Z_{v_2}$ and
denote by $w$ the only vertex of $V_1$ not chosen. Now we
denote by $Y_1$ the set $\{v_1\in V_1\ : \ d^+_{V_1}(v_1)=1\}$, and by
$X_1$ the set $N^+_{V_1}(Y_1)$.  If there exists $x_1\neq w$ in
$V_1\setminus X_1$, it means that every vertex in $V_1$ has an
out-neighbour in $V_1\setminus x_1$. Then we choose $y$ and $z$ in
$X_2=\{a,b,c\}$ such that $N^+_{V_2}(w)\not\subseteq
  \{y,z\}$. Now it is easy to check that the 2-colouring defined by
$(V_1\setminus \{x_1\})\cup \{y,z\}$ is a 2-out-colouring of $T$. So
we must have that $V_1\subseteq X_1\cup \{w\}$. In particular,
as $|X_1|=|Y_1|\le 3$, we have $|V_1|\le 4$. We will first look
at two particular cases. In {\bf case A} we assume that $T[X_1]$ is a
2-cycle or a 3-cycle  and that $w$ dominates all vertices of
  $X_1$. Then, we choose $x'\in X_1$, $x$ the out-neighbour of $x'$
in $T[X_1]$, $y$ the vertex of $V_2$ such that $x'\in Z_y$ and $z\in
X_2$. It is easy to check that $(V_1\setminus \{x\})\cup
\{y,z\}$ defines a 2-out-colouring of $T$. In {\bf case B} we assume
that $T[V_1]$ has size 3 and contains a 3-cycle $wx_1y_1$. We denote
by $x_2$ (resp. $y_2$) the only out-neighbour of $x_1$ (resp. $y_1$)
in $V_2$. If $N_{V_2}^+(w)=\{x_2\}$ then we check that $(V_1\setminus
\{w\})\cup \{y_2,a\}$ defines a 2-out-colouring of $T$. If
$N_{V_2}^+(w)\neq \{x_2\}$ then we can choose $z\in \{a,b,c\}$ such
that $N_{V_2}^+(w)\cap (V_2\setminus \{x_2,z\})\neq \emptyset$. So we
check that $(V_1\setminus \{y_1\})\cup \{x_2,z\}$ defines a
2-out-colouring of $T$.\\ Now, if $|V_1|=4$ then $X_1=Y_1=V_1\setminus
w$ and $T[V_1]$ is a 3-cycle dominated by $w$, situation we treated in
{\it case A}. If $|V_1|=3$ then either $|X_1|=2$ or $|X_1|=3$. In the
former case, if $Y_1=X_1$ then $T[V_1\setminus w]$ is a 2-cycle
dominated by $w$, corresponding to {\it case A}. If $|X_1|=2$
and $Y_1\neq X_1$ then $T[V_1]$ contains a 3-cycle, case we
already settled in {\it case B}. And if $|V_1|=|X_1|=3$ then $T[V_1]$
is exactly a 3-cycle, situation corresponding to {\it case B}. Finally
if $|V_1|=2$ then $T[V_1]$ is a 2-cycle. We denote by $wd$ this
2-cycle and by $e$ the only out-neighbour of $d$ in $V_2$. If $w$ has
at least one out-neighbour and at least one in-neighbour in $X_2$,
then we can assume that $w$ dominates $a$ and is dominated by $b$. So
$T$ has a balanced 2-out-colouring defined by $\{a,b,w\}$,
contradicting the choice of $\gamma$. Otherwise $X_2$ dominates
$w$ or is dominated by $w$ and there are no digons between $w$ and
$X_2$. In these case we can see that $T$ belongs to the family ${\cal T}_2$ or ${\cal T}_3$.  \qed

From the proof of the above theorem we can derive the following
corollary.

\begin{corollary}
There exists a polynomial algorithm which given a semicomplete digraph
$T\not\in {\cal T}_i$, $i\in [3]$ and a 2-out-colouring of $T$,
returns a balanced 2-out-colouring of $T$
\end{corollary}

\section{2-partitions of tournaments with out-degree at least $k$  to both sets}\label{kinbothsets}

In this section, for a given 2-partition $(V_1,V_2)$ we let
$\induce{D}{V_i}$ denote the induced subgraph of $D$ on $V_i$ and let
$\induce{D}{V_1,V_2}$ denote the spanning bipartite subgraph whose
edges are all edges of $D$ with an end vertex in $V_1$ and an end
vertex in $V_2$.  Recall that we call a 2-partition $(V_1,V_2)$
balanced if $||V_1|-|V_2|| \leq 1$.
\vspace{0.2cm}

Our aim is to use probabilistic methods to obtain results about
sufficient conditions, in terms of minimum out-degrees, for the
existence of 2-partitions where all vertices have at least $k$
out-neighbours in both sets.  Our main result is the following

\noindent
\begin{theorem}
\label{t12}
There exist two absolute positive constants $c_1,c_2$ so that the
following holds.
\begin{enumerate}
\item
Let $T=(V,E)$ be a tournament with minimum out-degree at least $2k+c_1
\sqrt k$.  Then there is a balanced partition $V=V_1 \cup V_2$ of $V$
so that $\delta^+(\induce{T}{V_1}), \delta^+(\induce{T}{V_2})$ and
$\delta^+(\induce{T}{V_1,V_2})$ are all at least $k$.
\item
For infinitely many values of $k$ there is a tournament with minimum
out-degree at least $2k+c_2 \sqrt k$ so that for any partition $V=V_1
\cup V_2$ of $V$ into two disjoint sets, at least one of the
quantities $\delta^+(\induce{T}{V_1}), \delta^+(\induce{T}{V_2})$,
$\delta^+(\induce{T}{V_1,V_2})$ is smaller than $k$.
\end{enumerate}
\end{theorem}

In order to illustrate the usefulness of the probabilistic method, we
first give short probabilistic proof of the following corollary of
Theorems \ref{theo:mindegree3} and \ref{theo:balanced}. Recall that
the constant 4 is best possible by the Paley tournament $P_7$.
  
\noindent
\begin{corollary}
\label{t11}
Let $T=(V,E)$ be a tournament with minimum out-degree at least
$4$. Then there is a balanced partition $V=V_1 \cup V_2$ of $V$ so
that $\delta^+(\induce{T}{V_1}), \delta^+(\induce{T}{V_2})$ and
$\delta^+(\induce{T}{V_1,V_2})$ are all at least $1$. 
\end{corollary}
\vspace{0.2cm}

\subsection{Proofs}

We need the following simple statement.
\begin{lemma}
\label{l21}
Let $(n_i)_{i \geq i_0}$ be a sequence of integers, put
$N_j=\sum_{i=i_0}^j n_j$ and let $(p_i)_{i \geq i_0}$ be a
non-increasing sequence of non-negative reals. Suppose $N_j \leq
B_j$ for all $j \geq i_0$. Then the sum 
$$
S=\sum_{i \geq i_0} n_i p_i 
$$
satisfies
$$
S \leq B_{i_0}p_{i_0} + 
\sum_{i \geq i_0} (B_{i+1}-B_i) p_{i+1}.
$$
\end{lemma}
\vspace{0.2cm}

\noindent
{\bf Proof:}\,   Let $S$ be as above, then
$$
S=N_{i_0}p_{i_0}+(N_{i_0+1}-N_{i_0})p_{i_0+1}
+(N_{i_0+2}-N_{i_0+1})p_{i_0+2}+ \ldots 
$$
$$
=N_{i_0}(p_{i_0}-p_{i_0+1})+N_{i_0+1}(p_{i_0+1}-p_{i_0+2}) + \ldots
$$
$$
\leq B_{i_0} (p_{i_0}-p_{i_0+1})+B_{i_0+1}(p_{i_0+1}-p_{i_0+2}) +
\ldots
=
B_{i_0}p_{i_0} +
\sum_{i \geq i_0} (B_{i+1}-B_i) p_{i+1},
$$
as needed. \qed

\vspace{0.2cm}

\noindent
{\bf Proof of Corollary \ref{t11}:}\, 
Let $n_i$ denote the number of vertices of $T$
with out-degree $i$, and put $N_j=\sum_{i \leq j} n_i$. 
Thus $N_3=0$. 
Note
that $N_i \leq 2i+1$ for all $i$, as the average out-degree 
in any induced subgraph of $T$ on more than $2i+1$ vertices
exceeds $i$. We consider two possible cases.
\vspace{0.1cm}

\noindent
{\bf Case 1:}\, $n_4 \geq 2$. Let $x,y$ be two vertices of
out-degree $4$ in $T$ and let $M$ be an arbitrary near perfect
matching in $T$ containing the edge connecting $x$ and $y$
and two additional  edges $x_1,x_2$ and $y_1,y_2$
where $x_1,x_2$ are out-neighbors of $x$ and $y_1,y_2$ are
out-neighbors of $y$ (it is easy to check that there are always
three disjoint edges as above).
For each edge $cd$ of $M$, randomly and
independently, place either $c$ in $V_1$ and $d$ in $V_2$ or $c$ in $V_2$
and $d$ in $V_1$, where each of these choices are equally likely.
If $|V|$ is odd place the remaining vertex, uncovered by $M$,
randomly and uniformly either in $V_1$ or in $V_2$. Note that
by construction $V_1$ and $V_2$ are of nearly equal sizes. In addition
both $x$ and $y$ have an out-neighbor in $V_1$ and an 
out-neighbor in $V_2$. Moreover, each of the vertices that is neither
an
out-neighbor of $x$ nor of $y$ has both $x$ and $y$ as out-neighbors,
and hence has at least one out-neighbor in $V_1$ and at least one in
$V_2$. The only remaining vertices are the out-neighbors of $x$ and
$y$ (besides  $x$ and $y$). There are at most $7$ such vertices.
For each such vertex $v$ let $A_v$ denote the event that $v$ fails
to have an out-neighbor in $V_1$ or in $V_2$. If the out-neighbors of
$v$ contain an edge of $M$, the probability of this event is $0$.
Else, its probability is exactly $2^{1-d^+(v)} \leq 1/8$, where 
$d^+(v)$ is the out-degree of $v$. Thus, by the union bound, the
probability that some event $A_v$ holds is at  most $7/8<1$,
showing that with positive probability no event $A_v$ holds, that
is, there is a partition with the desired properties in this case.
\vspace{0.2cm}

\noindent
{\bf Case 2:}\, $n_4 \leq 1$. Put $p_i=2^{1-i}$ and let $M$ be a
near perfect matching as before, including an edge
connecting two out-neighbors of the unique vertex of out-degree $4$,
if there is such a vertex. Let $V_1$ and $V_2$ be chosen randomly, as
before, by the random process splitting the endpoints of each edge
of $M$ randomly and independently between $V_1$ and $V_2$. Thus $V_1$ and
$V_2$ are of nearly equal sizes and if there is a vertex of degree
$4$ then it has out-neighbors in $V_1$ and in $V_2$. It remains to deal
with the other vertices. For each vertex $v$ let $A_v$ be the event
defined in Case 1. We have to show that with positive probability
no event $A_v$ holds. Using the fact that the sequence $p_i$ is
decreasing it follows, by Lemma \ref{l21},
that the sum of probabilities of all these
events  is at most
$$
11 p_5+\sum_{i \geq 5} ((2i+3)-(2i+1))p_{i+1}
=11 \cdot 2^{-4} +\sum_{i \geq 5} 2^{1-i} =13/16<1.
$$
This proves the existence of the desired partition.\\

We proceed with the proof of the second part of Theorem \ref{t12}. 
For a prime $q$ which is $3$ modulo $4$, the quadratic residue tournament
$P_q$ is the tournament whose 
vertices are the integers
modulo $q$ where $(i,j)$ is a directed edge iff $i-j$ is a
quadratic residue modulo $q$. 
\begin{lemma}
\label{l22}
Let $P_q=(V,E)$ be as above. Then for any function 
$f: V \mapsto \{-1,1\}$ there is a vertex $v \in V$
so that 
$ | \sum_{u \in N^+(v)} f(u) | > \frac{1}{2} \sqrt q$.
\end{lemma}
\vspace{0.2cm}

\noindent
{\bf Proof:}\, It is easy and well known 
(c.f., e.g., \cite{alon2016}, Chapter  9) 
that every vertex of $P_q$ has out-degree and in-degree $(q-1)/2$  
and any
two vertices of it have exactly 
$(q-3)/4$ common in-neighbors (and out-neighbors). 
Let $A=A_q$ be the adjacency matrix
of $P_q$, that is, the $0/1$ matrix whose rows and and columns are
indexed by the vertices of $P_q$, where $A_{ij}=1$ iff $(i,j)$ is a
directed edge. By the above comment each diagonal entry of
$A^tA$ is $(q-1)/2$ and each other entry is $(q-3)/4$. Thus the
eigenvalues of $A^tA$ are
$(q-1)/2+(q-1)(q-3)/4=(q-1)^2/4$ (with multiplicity $1$) and
$(q-1)/2-(q-3)/4=(q+1)/4$ (with multiplicity $(q-1)$). This implies that 
$$
||Af||_2^2 =f^t A^t A f \geq (q+1)/4 ||f||_2^2 =q(q+1)/4.
$$
It follows that there is an entry of $Af$ whose square is at least
$(q+1)/4$, completing the proof.  \qed

\vspace{2mm}

Note that by the above Lemma, for any partition of the vertices of
$P_q$ into two disjoint (not necessarily nearly equal) 
sets $V_1$ and $V_2$ there is a vertex $v$ of
$P_q$ so that the number of its out-neighbors in $V_1$ differs from
that in $V_2$ by more than $\sqrt q/4$. This implies the assertion of
part (ii) of Theorem \ref{t12} for infinitely many 
values of $k$.

Before proving the assertion of Theorem \ref{t12} we describe a
short proof of the following weaker result.
\begin{proposition}
\label{p23}
Let $T=(V,E)$ be a tournament with minimum out-degree at least
$$
2k+(1+o(1)) \sqrt {2 k \ln k}. 
$$
Then there is a balanced partition $V=V_1 \cup V_2$ of $V$ so that
$\delta^+(\induce{T}{V_1}), \delta^+(\induce{T}{V_2})$ and
$\delta^+(\induce{T}{V_1,V_2})$ are all at least $k$. (The $o(1)$-term
above tends to zero as $k$ tends to infinity.)
\end{proposition}
\vspace{0.1cm}

\noindent
{\bf Proof:}\, The proof is similar to that of Theorem \ref{t11}.  We
assume, whenever this is needed, that $k$ is sufficiently large.  Put
$m=2k+(1+\epsilon) \sqrt {2k \ln k}.$ Let $n_i$ denote the number of
vertices of $T$ with out-degree $i$, and put $N_j=\sum_{i \leq j}
n_i$.  Thus $N_s=0$ for all $s < m$ and $N_s \leq 2s+1$ for all $s$.
Let $M$ be an arbitrary near perfect matching in $T$.  For each edge
$cd$ of $M$, randomly and independently, place either $c$ in $V_1$ and
$d$ in $V_2$ or $c$ in $V_2$ and $d$ in $V_1$, where each of these
choices are equally likely.  If $|V|$ is odd place the remaining
vertex, uncovered by $M$, randomly and uniformly either in $V_1$ or in
$V_2$. Note that by construction $V_1$ and $V_2$ are of nearly equal
sizes.  For each vertex $v$ of $T$ let $A_v$ be the event that $v$ has
less than $k$ out-neighbors in $V_1$ or less than $k$ out-neighbors in
$V_2$. Let $d(\geq m)$ be the out-degree of $v$.  Note that $A_v$ is
exactly the event that the number of out-neighbors of $v$ in $V_1$
differs from that in $V_2$ by more than $d-2k$. Let $X_v$ be the
random variable whose value is this difference.  If the set of
out-neighbors of $v$ contains no edge of the matching $M$ then $X_v$
is the sum of $d$ independent uniform $-1,1$ variables. By the
Chernoff Inequality (c.f., e.g., \cite{alon2016}, Theorem A.1.2) the
probability of the event $A_v$ is at most $2e^{-(d-2k)^2/2d}$. If the
set of out-neighbors contains $t$ edges of the matching then $X_v$ is
the sum of only $d-2t$ independent uniform $-1,1$ variables and the
probability is even smaller. Put $p_d=2e^{-(d-2k)^2/2d}$. It is easy
to check that the sequence $p_d$ is decreasing for all $d \geq 2k$
(indeed the function $g(x)=-(x-2k)^2/2x$ is decreasing on the interval
$[2k,+\infty[$). Therefore, by Lemma \ref{l21} and the union bound,
    the probability that at least one of the events $A_v$ holds is at
    most
$$
P \leq 2 \cdot (2m+1)e^{-(m-2k)^2/2m} 
+2 \sum_{d >m} p_d.
$$
In the sum above each of the terms $p_d$ for $d \leq 4k$ is at most
$p_m$, and each of the terms $p_d$ for bigger $d$ is at most
$2e^{-d/8}$. Therefore
$$
P \leq (4m+2+2 \cdot 4k) e^{-(m-2k)^2/2m}  +2 \sum_{d >4k}
2e^{-d/8} < (1+o(1)) 16k e^{-(m-2k)^2/2m} +40 e^{-4k/8} <1,
$$
where in the last inequality we have used the assumption that $k$ is
large.
It follows that with positive probability none of the events 
$A_v$ holds, completing the proof. \qed

\vspace{2mm}

We proceed with the proof of part (i) of Theorem \ref{t12}. An
equivalent formulation of this part is that for any tournament
$T=(V,A)$ with
minimum out-degree at least $2k+c_1 \sqrt k$ there is a function
$f: V \mapsto \{-1,1\}$ so that $|\sum_{v \in V} f(v)| \leq 1$
and for every vertex $v$ of $T$ with out-degree $d=|N^+(v)|$
$$
|\sum_{u \in N^+(v)} f(u) | \leq d-2k.
$$
Indeed, given such an $f$ we can simply define $V_1=f^{-1}(1)$
and $V_2=f^{-1}(-1)$.  This resembles results about discrepancy of
set systems (see, e.g., \cite{alon2016}, Chapter 13). 
In particular, for the special case in which the number of
vertices of the tournament is at most, say, $10k$, the above
follows from the six standard deviations result of Spencer
\cite{spencerTAMS289}, which asserts that the discrepancy of any hypergraph
with $m$ vertices and $m$ edges is at most  $6 \sqrt m$. 
The general case requires some work, we prove it by combining a
variant of  the
partial coloring idea of \cite{spencerTAMS289} (see also \cite{beckC1}) 
with the main result of \cite{spencerTAMS289}. In what follows we make no
attempt to optimize the absolute constants.
\begin{lemma}
\label{l24}
Let $\FF$ be a family of subsets of $[n]=\{1,2,\ldots,n\}$, and
suppose that each  set $F \in \FF$ is of size at least
$2k+1000\sqrt k$. Suppose, further, that for every $s$ there are
less than $3s$ members of $\FF$ of size at most $s$. Then there is 
a function $f: [n] \mapsto \{-1,0,1\}$ such that
\begin{enumerate}
\item
$|\sum_{i=1}^n f(i)| \leq 1$ 
\item
For every $F \in \FF$ satisfying $|F| \leq 100k$,
$|\sum_{i \in F} f(i)| \leq 200 \sqrt k$
\item
For every $F \in \FF$ of size $|F|>100k$,
$|f^{-1}(1) \cap F| \geq k$ and
$|f^{-1}(-1) \cap F| \geq k$.
\end{enumerate}
\end{lemma}
{\bf Proof :}\, 
Without loss of generality assume that $n$ is even (otherwise  add
a point). The proof can be described using the pigeonhole
principle, but it is cleaner to present a version applying some
simple
properties of the entropy function. Recall that the (binary)
entropy of a random variable $X$ getting values $x_i$ with
probabilities $p_i$ for $i \in I$ is 
$$
H_2(X)=\sum_{i \in I} p_i \log_2 (1/p_i).
$$
It is well known (see, e.g., \cite{alon2016}, Chapter 15) that
if $X=(X_1,X_2, \ldots ,X_q)$ is a vector
then 
\begin{equation}
\label{e21}
H_2(X) \leq \sum_{i=1}^q H_2(X_i)
\end{equation}
Another fact we need is that 
\begin{equation}
\label{e22}
\mbox{If}~~ p_i \leq 2^{-r}~~ \mbox{for all}~~
i \in I~~\mbox{then} ~~H_2(X) \geq \sum_{i \in I} p_i \cdot r =r
\end{equation}
Let $g: [n] \mapsto \{-1,1\} $ be a random function
defined as follows. For each $i \leq n/2$ randomly, uniformly 
and independently put $g(2i-1)=1, g(2i)=-1$ or
$g(2i-1)=-1,g(2i)=1$.  Put $\FF_1=\{F \in \FF, |F| \leq 100k\}$.
For each $F \in \FF_1$ define 
$$
t(F)=\lfloor 0.5+\frac{\sum_{i \in F} g(i)}{20 \sqrt {|F|}} \rfloor
$$
Note that for each such $F$, $t(F)$ is very likely to be $0$. It is
$1$ or $-1$ only with probability smaller than $e^{-50}$, and more
generally it is $i$ or $-i$ with probability smaller than $e^{-50i^2}$
for all $i \geq 1$.  Therefore, the entropy of the random variable
$t(F)$ is smaller than
$$
\log_2 \frac{1}{1-2 e^{-50}} +\sum_{i \geq 1} 2 \cdot \log_2 e \cdot (50i^2)
e^{-50i^2}
$$
which is 
much smaller than $1/30000$.
Let $X$ be the (vector valued) random variable defined by
$X=(t(F): F \in \FF_1)$. By (\ref{e21}), since the total number
of members of $\FF_1$ is smaller than $300k$, the entropy of the 
vector $X$ is smaller than $0.01k$. It follows by (\ref{e22})
that there is a specific value of the vector $X$ obtained with
probability at least $2^{-0.01k}$. Fix such a vector $X$. In what
follows we show that we can choose a pair
$g_1,g_2$ of functions $g$ as above that give this vector
so that $f=(g_1-g_2)/2$ satisfies the assertion of the lemma.

We need the following claim.
\begin{claim}
\label{l25}
Let $F$ be a fixed member of $\FF -\FF_1$.  Then the number of pairs
of functions $g_1,g_2: [n] \mapsto \{-1,1\}$ defined in the previous
paragraph so that there are less than $k$ elements $j$ of $F$ with
$g_1(j)=1=-g_2(j)$ or that there are less than $k$ elements $j$ of $F$
with $g_1(j)=-1=-g_2(j)$ is smaller than $2^{n-0.1|F|}$.
\end{claim}
\vspace{0.1cm}

\noindent
{\bf Proof of Claim \ref{l25}:}\, 
Let $s$ be the
number of indices $i$ so that $\{2i-1,2i\} \subset F$.  Then there
are $|F|-2s$ elements of $F$ whose mate in the matching
$\{2i-1,2i\}, (i \leq n/2)$ is not in  $F$. Thus there are
exactly $2^{s+|F|-2s}=2^{|F|-s}$ ways to choose the values of
$g_1(j)$ for all $j \in F$ and the same number of ways to choose
the values of $g_2(j)$ for $j \in F$.  We next bound the number 
of choices in which there are less than $k$ elements $j$ of $F$
satisfying $g_1(j)=1=-g_2(j)$. This number is at most
\begin{equation}
\label{e23}
M=\sum_{j +\ell< k}{s \choose j} 2^s {{|F|-2s} \choose \ell}  
3^{|F|-2s-\ell} <{{|F|} \choose k} 2^s 3^{|F|-2s}.
\end{equation}
Indeed, for each $j, \ell$ with $j +\ell <k$ there are 
${s \choose j}$ ways to choose $j$ pairs $\{2i-1,2i\}$ contained in 
$F$ in which $g_1,g_2$ do not agree. Once these are chosen there
are still $2^s$ ways to choose the actual values of $g_1,g_2$ on
these $s$ pairs (agreeing on $j$ pairs and disagreeing  on $s-j$). 
There are then ${{|F|-2s} \choose \ell}$ ways to select  the 
elements $j \in F$ that do not belong to these pairs for which
$g_1(j) =1=-g_2(j)$. Finally there are $3$ possibilities
for the values of $g_1(j)$ and $g_2(j)$ for each other element $j$. 

Since $|F| \geq 100k$ it follows that 
${{|F|} \choose k} \leq 2^{H_2(0.01)|F| } < 2^{0.1|F|}$.
Here we used the known fact that for every pair of integers
$a>b>0$, ${a \choose b} \leq 2^{H_2(b/a) a}$ where
$$
H_2(x)=x \log_2 \frac{1}{x} +(1-x) \log_2 \frac{1}{1-x}
$$ 
is the binary entropy
of the number $x$, $0 <x<1$, which is the binary entropy of the
indicator random variable attaining the value $1$ with probability
$x$ and the value $0$ with probability $1-x$. The above inequality
follows, for example, from the assertion of Corollary 15.7.3 in
\cite{alon2016} by taking $n=a$ with $\FF$ being the family of all
subsets of cardinality $b$ of $\{1,2,\ldots ,a\}$.

Plugging in (\ref{e23}) we conclude that 
$$
M \leq 2^{0.1|F|} 2^s 3^{|F|-2s} < 2^{0.1|F|} 4^{|F|-s}
(3/4)^{|F|} <2^{-0.3|F|} 4^{|F|-s}<
2^{-0.11|F|} 4^{|F|-s}
$$
Each choice of the values of $g_1,g_2$ on the $|F|-s$ pairs of
elements  $\{2i-1,2i\}$ intersecting $F$ can be completed to the
full values of $g_1,g_2$  in $4^{n/2-|F|+s}$ ways. Thus the number
of pairs $g_1,g_2$ for which there are less than $k$ elements $j$
with $g_1(j)=1=-g_2(j)$ is smaller than
$$
2^{-0.11|F|} 4^{|F|-s} \cdot 4^{n/2-|F|+s}=
2^{n-0.11|F|}
$$
By symmetry, the number of pairs $g_1,g_2$ so that 
there are less than $k$ elements $j$
with $g_1(j)=-1=-g_2(j)$ satisfies the same inequality. This
provides the assertion of Claim~\ref{l25}. \qed

\vspace{0.2cm}

\noindent
Returning to the proof of Lemma \ref{l24} recall that we have fixed
a value of $X$ such that there are at least $2^{n/2-0.01k}$
choices for the function  $g$ giving the value of $X$. Hence
there are at least $2^{n-0.02k}$ choices for an ordered pair of
functions $g_1,g_2$ giving this value. By Claim~\ref{l25}, among
these choices, the number of pairs that have, in some
set $F \in \FF-\FF_1$, less than $k$ elements
$j$ with $g_1(j)=1=-g_2(j)$ or less than $k$ elements
$j$ with $g_2(j)=1=-g_1(j)$ is at most
$$
\sum_{F \in \FF-\FF_1} 2^{n -0.1 |F|}
< \sum_{s \geq 100k} (3s) 2^{n-0.1s} < 2^{n-0.02k}
$$
(with a lot of room to spare). 

In the above inequality we used the fact that for
$A=\sum_{s=r}^{\infty} sq^s$, $qA=\sum_{s=r}^{\infty} sq^{s+1}$, and
thus
$$
A-qA=rq^r +\sum_{s=r+1}^{\infty}  q^s =rq^r +\frac{q^{r+1}}{1-q}
$$
So we obtain
$$
A=\frac {rq^r}{1-q} + \frac{q^{r+1}}{(1-q)^2}
$$
Taking $q=2^{-0.1}$ and $r=100k$ gives the required bound. 

Since there are $2^{n-0.02k}$ choices for the ordered pair of
functions $g_1,g_2$, each giving the value of $X$, it follows
that there is a pair $g_1,g_2$ for which the event
described in Claim~\ref{l25} 
does not happen for any $F \in \FF-\FF_1$. Fix such $g_1,g_2$ and
define $f=(g_1-g_2)/2$. We claim that $f$ satisfies the assertion
of Lemma~\ref{l24}. Indeed, by construction it satisfies property
1. Property 2 follows from the fact that both $g_1$ and $g_2$
give the same vector $X$ implying that for every
$F \in \FF_1$
$$
\lfloor 0.5+\frac{\sum_{i \in F} g_1(i)}{10 \sqrt {|F|}} \rfloor
=
\lfloor 0.5+ \frac{\sum_{i \in F} g_2(i)}{10 \sqrt {|F|}} \rfloor
$$
Property 3 follows from the fixed choice of the functions
$g_1,g_2$. This completes the proof of the lemma.
\qed
\vspace{0.2cm}

\noindent
{\bf Proof of Theorem \ref{t12}, part (i):}\, 
Let $[n]=\{1,2,\ldots ,n\}$
denote the set of vertices of $T$ and let
$\FF$ be the family of out-neighborhoods of its vertices.
Note that it satisfies the assumptions of Lemma~\ref{l24}.
Let $f$ be as in the conclusion of the lemma. Put all
elements in $f^{-1}(1)$ in $V_1$ and all elements in
$f^{-1}(-1)$ in $V_2$ (the partition of the elements in $f^{-1}(0)$
will be determined in what follows.) Note that the partial
assignment  above already ensures, by property 3, that all vertices with 
out-degree exceeding $100k$ have at least $k$ out-neighbors in 
$V_1$ as well as in $V_2$.

It remains to assign the vertices $i$ for which $f(i)=0$ values in
$\{-1,1\}$ ensuring that we do not increase the discrepancy of the
members of $\FF_1$ by too much and maintaining the property $|\sum
f(i)| \leq 1.$ We then define $V_1=f^{-1}(1), V_2=f^{-1}(-1)$.  This
is done by applying the result of \cite{spencerTAMS289} that implies
that the discrepancy of any set system of $m$ sets (of any sizes) is
at most $12 \sqrt m$, (see \cite{alon2016}, Corollary 13.3.4).  Apply
that to the family consisting of the intersections of the sets in
$\FF_1$ with $f^{-1}(0)$ together with one additional set: the set
$f^{-1}(0)$. Adding this set ensures that our resulting partition is
nearly balanced, and we can now change the values of some $O(\sqrt k)$
elements (among those in $f^{-1}(0)$) arbitrarily, to make the split
precisely balanced without changing the discrepancy of any set $F$ in
$\FF_1$ by more than $2 \cdot 12 \sqrt {300k +1}$ (an addition of at
most $12 \sqrt {300k +1}$ corresponding the discrepancy of $F \cap
f^{-1}(0)$ and another addition of at most this quantity corresponding
to the final arbitrary modification of values that makes the split
balanced).  This completes the proof.  \qed

\subsection{Consequences of our methods}We conclude this section with some observations about further consequences of the methods and proofs we used.
\begin{itemize}
\item The proofs carry over with no change to semicomplete digraphs.
\item The same proof works to split a tournament or a semicomplete
  digraph to $r$ parts with each vertex having at least $k$
  out-neighbors in each part. Minimum out-degree $(1+o(1))rk$
  suffices.  Getting best possible bounds for small $r$ and $k$, e.g.,
  $r \leq 10$ and $k=1$, may be difficult.
\item
The same proof works to split an oriented graph with no independent set
of size $s$. Here any set of size exceeding $(s-1)(2r+1)$ contains 
a vertex of out-degree exceeding $r$, hence we have an upper bound
on the number of vertices of out-degree at most $r$ and can repeat
the probabilistic argument to get a similar result. 
\item
The probabilistic proof, with no real essential change,
can handle simultaneously 
in-degrees and out-degrees, establishing the following.
\begin{theorem}
\label{t31}
There exists an absolute constant $c$ so that the
following holds for every positive integer $k$. 
Let $T=(V,A)$ be a tournament with minimum
out-degree at least $2k+c \sqrt k$ and minimum in-degree at least
$2 k + c \sqrt k$.
Then there is a balanced partition $V=V_1 \cup V_2$ of
$V$  so that every vertex has at least
$k$ out-neighbors in $V_1$ and in 
$V_2$, and every vertex has at least $k$ in-neighbors 
in $V_1$ and in $V_2$.
\end{theorem}

\end{itemize}

\section{Further remarks}\label{remarksec}

We have shown in Section~\ref{kinbothsets}, using probabilistic
arguments, that every tournament whose minimum out-degree is
sufficiently large as a function of k has a partition of the vertices
into two sets of nearly equal sizes so that each vertex has at least k
outneighbours in each set. The bound we give on this function is
asymptotically best possible.  After the completion of this paper we
learned from the authors of~\cite{yangArXiv17} that they have
independently proved the existence of such a function. Their proof is
also probabilistic, but the bound they get for the function is weaker,
roughly twice our bound.\\

A digraph is $k$-out-regular is all out-degrees are $k$.  As mentioned
in the introduction Thomassen \cite{thomassenEJC6} showed that for
every integer $k\geq 1$ there exist $k$-out-regular digraphs with no
even cycle. As mentioned in \cite{alonCPC15} this implies that for
every $k$ there is an oriented graph with minimum out-degree at least
$k$ which admits no splitting into two parts with every vertex having
an out-neighbor in the other part. On the other hand, as shown here
and as mentioned above, such splittings always exist for tournaments
and for graphs with bounded independence numbers, provided the minimum
out-degree is large enough.  It may be interesting to give additional
natural classes of oriented graphs for which such a splitting is
possible.




\begin{proposition}
There exists a polynomial algorithm for deciding whether a
2-out-regular digraph has a 2-out-colouring.
\end{proposition}

\pf We define the non-oriented graph $G_D$ on $V(D)$ with edge set
$\{N^+(x)\ :\ x\in V(D)\}$.  The result follows from the fact that a
2-out-regular digraph $D$ has a 2-out-colouring if and only if the
graph $G_D$ defined above is bipartite. This can be checked in linear
time. \qed.

As mentioned in the introduction, there is a polynomial algorithm for
deciding whether a digraph $D$ has a 2-partition $(V_1,V_2)$ such that
each of $\induce{D}{V_i}$, $i=1,2$ have out-degree at least one. It
was shown in \cite{bangman17} that, despite Thomassen's examples
mentioned above, it is also polynomial to decide whether $D$ has a
2-partition so that $\induce{D}{V_1,V_2}$ has minimum out-degree at
least 1. So, given that 2-out-colouring is ${\cal NP}$-complete, it is
natural to ask about the complexity of deciding whether $D$ has a
2-partition such that $D(V_1,V_2)$ and $\induce{D}{V_1}$ both have
out-degree at least 1, but we do not require this for
$\induce{D}{V_2}$. The following result shows that this is also ${\cal
  NP}$-complete.

\begin{theorem}
It is ${\cal NP}$-complete to decide whether a given digraph $D$ has a
2-partition $(V_1,V_2)$ so that $D(V_1,V_2)$ and $\induce{D}{V_1}$
both have minimum out-degree at least one.
\end{theorem}

\pf Call a 2-partition $(V_1,V_2)$ {\bf nice} if $D(V_1,V_2)$ and
$\induce{D}{V_1}$ both have minimum out-degree at least one. Let $X$
denote the digraph on 6 vertices $\{a,b,v,v',\bar{v},\bar{v}'\}$ and
the following 12
arcs\\ $\{ab,ba,av',b\bar{v}',vv',v'v,\bar{v}\bar{v}',\bar{v}'\bar{v},v\bar{v},
\bar{v}v', v'\bar{v}',\bar{v}'v\}$. It is not difficult to check that
$X$ has a nice 2-partition and that for every such partition
$(V_1,V_2)$ we have $v,v'\in V_1$ and $\bar{v},\bar{v}'\in V_{3-i}$
for $i=1$ or $i=2$ and both are possible. Now let ${\cal F}$ be an
instance of NAE-3-SAT with variables $x_1,x_2,\ldots{},x_n$ and
clauses $C_1,C_2,\ldots{},C_m$. Form a digraph $R=R({\cal F})$ as
follows: take $n$ copies $X_1,X_2,\ldots{},X_n$ of $X$ and give the
vertices in the $i$th copy subscript $i$, i.e. in $X_i$ the vertex
$v_i$ corresponds to the vertex $v$. Associate the vertices
$v_i,\bar{v}_i$ of $X_i$ with the variable $x_i$ and its negation
$\bar{x}_i$, respectively. Now for each clause $C_j$ we add two new
vertices $d_j,c_j$, the arc $d_jc_j$ and three arcs from $c_j$ to
those vertices in the $X$ copies that correspond to the literals of
$C_j$. Hence if $C_j=(x_4\vee{}\bar{x}_6\vee{} x_8\}$, then we add the
three arcs $c_jv_4,c_j\bar{v}_6$ and $c_jv_8$. This completes the
description of $R$. As the vertices $d_j$, $j\in [m]$ have out-degree
one, they must all belong to $V_2$ in any nice 2-partition and this
forces all the vertices $c_1,c_2,\ldots c_m$ to belong to $V_1$ for
all nice 2-partitions. Assume now that $R$ has a nice 2-partition
$(V_1,V_2)$.  By the remark above, each vertex $c_j$ belongs to $V_1$
and hence must have an out-neighbour in both sets and it is easy to check
that if we set $x_i$ to be true whenever $v_i\in V_1$ and false
otherwise, we obtain a truth assignment satisfying one or two literals
of every clause. Conversely, given a truth assignment $\phi$ satisfying
at least one but never three literals of any clause, we obtain a nice
2-partition by putting $\{d_1,\ldots{},d_m\}$ in $V_2$,
$\{c_1,c_2,\ldots{},c_m\}$ in $V_1$, for each $i\in [n]$ putting
$v_i\in V_1$ and $\bar{v}_i\in V_2$ if $\phi(x_i)='True'$ and $v_i\in
V_2$ and $\bar{v}_i\in V_1$ if $\phi(x_i)='False'$ and finally
distributing the rest of the vertices of the $X$ copies in $V_1,V_2$
(as we know we can). \qed

\bibliography{refs}

\end{document}